\documentclass[a4]{aa}
\usepackage[dvips]{graphics}

\begin{document}

\title{First results from the ESO VLTI Calibrators Program
\thanks{Table~4 is only available
 in electronic form
at the CDS via anonymous ftp to cdsarc.u-strasbg.fr (130.79.128.5)
or via http://cdsweb.u-strasbg.fr/cgi-bin/qcat?J/A+A/
}
}

\author{
        A. Richichi%\inst(1)
    \and
        I. Percheron%\inst{1}
}
\offprints{A. Richichi}

\institute{European Southern Observatory,
Karl-Schwarzschildstr. 2, D-85748 Garching bei M\"unchen, Germany\\
\email{arichich@eso.org}
}

\date{Received / accepted }

\abstract{
The ESO Very Large Telescope Interferometer (VLTI) is one of the
leading interferometric facilities.
It is equipped with several
8.2 and 1.8\,m telescopes, a large number of baselines up to 200\,m,
and with several
subsystems designed to enable high quality
measurements and to improve significantly 
the limits of sensitivities currently
available to long-baseline interferometry. The
full scientific potential of the VLTI can be exploited only
if a consistent set of good quality calibrators is available.
For this, a large number of observations of potential
calibrators have been obtained during the commissioning phase
of the VLTI. These data are publicly available.
We briefly describe
the interferometer, the VINCI instrument used for the observations,
the data flow from acquisition to processed results, and we
present and comment on the volume of observations gathered and
scrutinized. The result is a list 
191 calibrator candidates, for which a total 
of 12066 observations can be deemed of satisfactory quality.
We present a general
statistical analysis of this sample, using as a starting point
the angular diameters previously available in the literature.
We derive the general characteristics of
the VLTI transfer function, and its
trend with time in the period 2001 through mid-2004.
A second paper will be devoted to a detailed investigation
of a selected sample, aimed at establishing a VLTI-based
homogeneous system of calibrators.
 }
\maketitle

\section{Introduction}\label{introduction}
Long-baseline interferometers provide at present
the highest angular resolution possible from
the ground at optical and infrared wavelengths. 
In the last few years, the CHARA, 
Keck and VLTI  interferometers
(ten Brummelaar et al. \cite{chara}, Colavita \& Wizinowich
\cite{keck}, Glindemann et al. \cite{Gli03})
have started operation, while other facilities such as
OHANA (Perrin et al. \cite{ohana}) and MRO (Creech-Eakman et al. 
\cite{mro}) are
being readied.
These new large interferometers couple for
the first time hectometric baselines with large telescopes,
offering a wavelength coverage from the visual
to the thermal infrared and aiming also at nulling interferometry.

Ground-based long-baseline interferometry (LBI)
is a technique that is heavily affected by atmospheric turbulence.
This sets limits on the integration times,
on the quality of the wavefronts that are combined from the various
telescopes, and in general on the accuracy of the measurements.
Although significant improvements can be obtained by selecting
observational sites of excellent quality and by adopting techniques
such as adaptive optics, fringe tracking and beam filtering
through monomode fibers,
the process of calibration remains a
fundamental requisite for the success of interferometric
measurements, and this aspect has only become more pressing
in view of the needs in terms of accuracy
demanded by the most recent interferometers
mentioned above. Indeed, the accuracy of calibration is at present
the limiting factor in measurements performed with the
VLTI, which additionally is located in the southern
hemisphere  where interferometric calibrators have been
traditionally more scarce.

The problem of calibration, although crucial for the
success of interferometric observations, has remained
until now a relatively uncoordinated effort.
In general, the tendency has been to use different, often
independent lists of calibrators at each interferometric
facility.
Another approach has been to estimate the angular diameter
for a set of suitable stars. Cohen et al.
(\cite{Coh99}) have produced a list of over 400 sources,
with typical magnitudes and diameters which are well suited
mostly for baselines of less than 100\,m and relatively
small telescopes. Bord{\'e} et al. (\cite{Bor02}) have
refined this list, imposing selection criteria
derived from their experience at the IOTA interferometer,
while M{\'e}rand et al. (\cite{Mer04}) have extended it
to stars with smaller angular diameter and suitable for
hectometric baselines.

The recent CHARM2 catalogue (Richichi et al. \cite{CHARM2})
is an effort to include all known direct measurements at high
angular resolution of angular diameters and binary stars, as well
as indirect estimates such as those mentioned above.
However, the best approach to the problem of calibrators,
at least those that can be resolved by interferometric
facilities, is undoubtedly represented by direct observations.

In this paper, we present the first results of an effort
to establish a network
of calibrators for the VLTI. We review the main
characteristics of the interferometer and its instrumentation,
and the observations carried out as part of the VLTI
commissioning.
We also briefly describe the VINCI
instrument used for such observations, and the process of
automated data reduction employed.
We then present the list of observed objects
and the main statistical conclusions, including an assessment
of the performance of the VLTI, its accuracy and stability.
A detailed analysis and discussion of individual
results is the subject of a second paper.

\section{VLTI overview}\label{VLTI}
The ESO Very Large Telescope Interferometer (VLTI), located on Cerro Paranal
in Chile, is based on the same site and infrastructure as the widely
known VLT observatory (Glindemann et al. \cite{Gli00}).
In addition to the four 8.2\,m unit telescopes
(UT), it includes a number of 1.8\,m auxiliary telescopes (AT).
These are expecially designed for interferometry, with very compact
structures and an optical system which replicates the light path
inside the UT telescopes (Koehler \& Flebus \cite{Koe00}).
Currently, one AT is available, two more are being delivered
and a fourth one is in an advanced phase of construction.
Also, two 0.4\,m siderostat test telescopes (SID,
Derie et al. \cite{Der00a}) are available, which have represented
the workhorse of the commissioning activities of the VLTI
(Sch{\"o}ller et al. \cite{schoeller}).
The ATs and SIDs can be moved over an array of 30 stations. In particular,
the ATs can be moved semi-automatically on a system of tracks and can be
relocated in a matter of hours; this is a unique feature among
large optical interferometers. Altogether,
254 independent baselines are available, leading to 3025 closure phases,
with a rather homogenous distribution in baseline lengths (from 8 to 205\,m) and orientations
(Richichi et al. \cite{Ric00}).

The VLTI is following a dense implementation plan, with several subsystems
and instruments
becoming available and being commissioned in close sequence
(Sch{\"o}ller et al. \cite{schoeller}).
Table~\ref{VLTI_schedule} provides
an overview of the main steps. More details
can be found in Derie (\cite{Der00b}) for the delay lines, 
Gai et al. (\cite{Gai03})
for FINITO, Donaldson et al. (\cite{Don00}) for MACAO,
Koehler \& Flebus (\cite{Koe00}) for the ATs, Derie et al. (\cite{Der03})
for PRIMA.

\begin{table}[h]
\caption{Schedule of VLTI subsystems and instruments.}
\label{VLTI_schedule}
\begin{tabular}{lll}
\hline
\hline
System & \multicolumn{1}{c}{Description} & \multicolumn{1}{c}{Status (Oct 04)}\\
\hline
VINCI  & Test instrument & available\\
MIDI  & $10 \mu$m instrument & available \\
DL 1-6  & Six Delay Lines & available \\
AMBER  & $1-2.4 \mu$m instrument & commissioning \\
FINITO & IR Fringe tracker & commissioning \\
MACAO 1-3  & UT Adaptive Optics & available\\
MACAO 4  & UT Adaptive Optics & integration\\
AT 1  & Auxiliary telescope & available \\
AT 2  & Auxiliary telescope & integration\\
AT 3\&4  & Auxiliary telescopes & Europe\\
SID & Test Siderostats & to be decomm.\\
PRIMA  & Dual-feed \& phase & from 2006 \\
& \multicolumn{1}{r}{referencing} & \\
\hline
\hline
\end{tabular}
\end{table}

The VLTI instruments and their characteristics are crucial
in defining the science capabilities of the interferometer,
and the specifications on the calibrators.
The first of these instruments is VINCI, developed by the
Paris Observatory in collaboration with ESO (Kervella et al. \cite{Ker00}). 
It achieved
first fringes with the SIDs in March 2001, and with the UTs in October 2001.
VINCI operates with a broad K-band filter
using a Hawaii array as detector. 
For most of its operation, VINCI has been used with 
the fiber-based beam combiner MONA (see Sect.~\ref{vinci}).
Note that at the time of writing another beam combiner is being
used with VINCI, based on integrated optics
(IONIC, 
Laurent et al. \cite{Lau02},
Le Bouquin et al. \cite{Lebo04}).
IONIC is available in two versions, that cover the H and 
the K bands respectively.
No extensive calibrator observations have been collected and made
public yet
with IONIC, and we will restrict ourselves 
in this paper
to VINCI operation
in 
a broad-band
K filter
with the MONA beam-combiner.

While VINCI was designed and built mainly to test and commission the
VLTI, the two facility instruments 
are MIDI and AMBER, both of which are installed 
and operating at the VLTI.
In particular, MIDI (Leinert et al. \cite{Lei00}) operates
in the N band, with a foreseen extension to the Q band.
It offers various levels of
spectral resolution and it represents one of very few
instruments of its kind in the world, and the only
one in the southern hemisphere.
AMBER (Petrov et al. \cite{Pet00}) is based
on a concept of a fiber-based beam combination similar to
VINCI. It offers 3 bands in the near-infrared
with various levels of spectral resolution, and the possibility
to combine either 2 or 3 beams. In order to achieve the
highest possibile accuracy on differential phase
measurements demanded by some of its scientific goals,
AMBER employs also a scheme of rapid beam switching.

\begin{table}[h]
\caption{Summary of VLTI first generation instruments and their
characteristics related to the needs of calibration.}
\label{VLTI_inst}
\begin{tabular}{lccc}
\hline
\hline
System & VINCI & MIDI & AMBER \\
\hline
Bands           & K & N  & J,H,K \\
$\lambda\lambda$ ($\mu$m)   & 1.9-2.4 & 8-13 & 1.0-2.4 \\
Ang. Resol.$^a$ (8\,m)      & 38.7 & 182-364   & 22.7-40.0 \\
Ang. Resol.$^a$ (202\,m)    & 1.5 &   7.2-14.4 &  0.9-1.6 \\
Spectr. Resol. ($\lambda / \Delta\lambda$) & 5 & 30-260 & 35-14000 \\
Beams           & 2 & 2 & 3 \\
Diff. phase         & no & no & yes \\
Polarization        & no & no & no  \\
Accuracy ($\Delta$V/V)  & 10$^{-3}$ & 10$^{-2}$ & 10$^{-4}$ \\
Lim. Magnitude     & 10$^b$ & 4$^c$ & 13$^d$ \\
\hline
\hline
\end{tabular}

$a$) Expressed in $0\farcs001$ for 50\% visibility at the
central wavelengths of the shortest and longest band, when
applicable. \\
$b$) effectively achieved during commissioning, in K-band,
using MACAO.
\\
$c$) 
N-band magnitude 
(goal)
for operation with the FINITO fringe tracker. 
From Leinert et al. (\cite{Lei03}).\\
$d$) K-band magnitude 
(goal)
for 5$\sigma$ fringe detection in 100\,ms on UTs equipped
with MACAO (Strehl=0.5). From Petrov et al. (\cite{Pet01}).

\end{table}

The main characteristic of the VLTI instruments are summarized
in Table~\ref{VLTI_inst}. 
We stress that in the case of the AMBER and MIDI instruments,
which are still in a preliminary phase of operation
without the full VLTI or not entirely within specifications,
the values listed are relative to the foreseen performance only.
Of particular interest are the
angular resolution, the sensitivity limits and the accuracy
of the visibility measurements. These represent the main parameters
required for the selection of
VLTI calibrators.
However, a number of additional
steps are currently under study, that could change significantly
the scenario provided so far. For example, the dual-feed PRIMA
facility (Derie et al. \cite{Der03}) will push by several magnitudes
the sensitivity limits, in those cases in which a nearby bright
star could be used for fringe tracking. It is foreseen that
the limiting sensitivity in K band would then reach about
20\,mag about 4 hours of integration. 
ESO and the European Space Agency ESA are currently
studying the possibility to build GENIE, a ground demonstrator
of the DARWIN nulling interferometer that will operate
later in space (Fridlund et al. \cite{Fri03}). Finally, a number
of concepts for a second generation of VLTI instruments
are also being considered (Glindemann et al. \cite{Gli03}).
For the immediate purpose of this paper, we will not
discuss the implications of such developments, that
would require a new set of calibrators.

\section{VINCI/MONA observations and data analysis}\label{vinci}
\begin{figure}
\resizebox{\hsize}{!}{\includegraphics{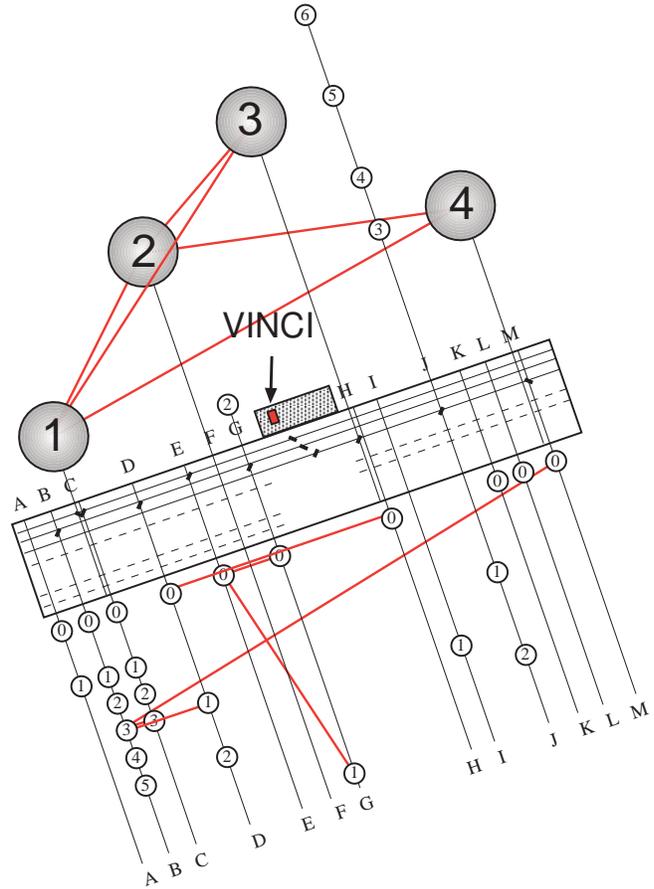}}
\caption[ ]{
Schematic layout of the VLTI interferometer, showing the
baselines used for the calibrators observations as listed in
Table~\ref{table_summary}. North is up.
}\label{fig_layout}
\end{figure}

The general layout of the VLTI system is shown in Fig.~\ref{fig_layout}.
Two afocal beams of light from either two siderostats or 
two UTs are reduced in diameter and sent through
an underground light duct system to the interferometric tunnel where
they are compensated for optical path differences by delay lines.
The beams are then directed into the interferometric laboratory.
Here, they are compressed in diameter from 80 to 18\,mm
(note this will not be required for AT operation),
and are
sent to the VINCI instrument where
they are focused onto single-mode optical fibers and afterwords
combined by a fiber beam combiner providing two interferometric outputs.
Before beam combination, the photometric signals are separated from each
of the two input beams in order to monitor the photometric fluctuations.
This design allows the higher order modes of the incoming radiation to 
be filtered and achieves interference with beams which are only affected 
by the lower order modes, specifically by tip-tilt induced by atmospheric 
piston noise.
This permits optimal performance
in terms of accuracy of visibility measurements. 
Thanks also
to the stability of the VLTI environment, VINCI attains often
uncalibrated visibility accuracies at the 0.1\% level for many hours
during nights of good seeing. 
This potential accuracy
is often not fully exploited, due
to the lack of suitable calibrators.

The two interferometric and two photometric output signals are imaged
simultaneously on the detector.
The fringes are detected by modulation of the optical
path difference with an amplitude of typically
200\,$\mu$m to 300\,$\mu$m using a piezo element within the VINCI
instrument, and kept within the scan length by sending an offset
signal to the delay lines at slow rates of about 1~Hz.
The two interferometric outputs, which are in opposite phase,
are subtracted to get a single interferogram which is free of
systematic system noise.
The squared coherence factor is obtained for each interferogram.
A detailed description of the algorithms used in processing
the VINCI data is given by Kervella et al. (\cite{Ker04}).
Note that two main approaches can be selected, based on
Fourier and wavelet transforms respectively. Experience
in the VLTI group with repeated measurements on the same source
and with large volumes of data has shown that the two approaches
lead essentially to the same results and accuracy, although
the wavelet analysis is more accurate on single measurements. 
We have used this latter for our computations of the transfer
function presented in Sect.~\ref{observations}.
An observation consists of the execution of an observing block (OB),
which is a sequence of typically 300 to 500 scans.
This sequence is also called a batch.

A description of the VLTI data flow, in connection also with VINCI,
can be found in Ballester et al. (\cite{Bal02}, \cite{Bal04}).
In summary, two data reduction systems are available.
Firstly, 
an automated data reduction through 
the so-called pipeline is triggered by the execution
of an OB on Paranal. A similar pipeline can be
initiated manually off-line with
parameter options in Garching
as we did.  This produces a synthetic ASCII summary
called the quality control (QC) log.
QC logs include among others the source name, time of observation,
baseline, telescope type, number of scans utilized,
the coherence factor and errors of each interferogram
as well as their average, and quality flags.

Secondly, various data reduction packages are available,
some of which are based on the same core of algorithms as the pipeline 
with an added library of graphical interfaces and scan selection routines,
and some of which start from the data produced by the pipeline
and provide additional selection and fitting algorithms.
This interactive data reduction is used for an accurate analysis
of individual observations. However, for the large volume of 
calibrator observations discussed here this approach is not practical.
We reprocessed
all the raw data relevant to this paper
with a single homogeneous 
version of the pipeline installed at the ESO headquarters in Garching,
and we used the
average values present in the 
QC logs.  

Note that the pipeline applies flags to
filter out individual scans and full OBs 
which do not meet a number of criteria. 
The settings that we used were such that a scan was rejected
if the signal-to-noise ratio (SNR)  on the two photometric
beams differed by more than a factor of 5; if the
fringe packet was too wide or too narrow in the time domain (more than
50\% away from the theoretical value); if the 
fringe peak was not at the correct position in the frequency domain (more than
30\% difference from the theoretical position); if the
fringe peak was too wide or too narrow in the frequency domain (more than
40\% difference from the theoretical value); if there was an optical path
difference (OPD) jump of more than 20$\mu$m before the scan; if
fringes were located at edge of the scan; and if the scan visibility
was more than 3$\sigma$ from the OB average.
A full OB was rejected if
less than 10 interferograms could be successfully processed; if
less than 5\% of the scans could be successfully  processed; if
the photometric signal averaged SNR less than 1.
More details on these criteria can be found in the paper
by Kervella et al. (\cite{Ker04}).

\begin{table}[h]
\caption{Statistics of the VINCI/MONA observations}
\label{table_summary}
\begin{tabular}{lrrrr}
\hline
\multicolumn{1}{c}{Baseline}& 
\multicolumn{1}{c}{Length}& 
\multicolumn{1}{c}{Nights}& 
\multicolumn{1}{c}{Total}& 
\multicolumn{1}{c}{Accepted}\\
&
\multicolumn{1}{c}{(m)}& 
&
\multicolumn{1}{c}{\# OB}& 
\multicolumn{1}{c}{Cal. \# OB}\\
\hline
B3-C3	&	8.0	&27	&	860	&426	\\
B3-D1	&	24.0	&63	&	2675	&1667	\\
B3-M0	&	139.7	&81	&	1856	&1118	\\
D0-H0	&	64.0	&50	&	825	&690	\\
E0-G0	&	16.0	&294	&	8317	&5606	\\
E0-G1	&	66.0	&145	&	3147	&2188	\\
U1-U2	&	56.6	&1	&	8	&11\\
U1-U3	&	102.4	&15	&	422	&250	\\
U1-U4	&	130.2	&1	&	41	&9	\\
U2-U3	&	46.6	&9	&	149	&85	\\
U2-U4	&	89.4	&2	&	52	&16	\\
\hline
Total	&		&	688	&	18352	&	12066	\\
\hline
\hline
\end{tabular}
\end{table}

A summary of the total
observations, ordered by baseline, 
is given in Table~\ref{table_summary}.
The last two columns list for each
baseline, respectively, 
the total number of OBs succesfully processed
and the number of candidate
calibrator OBs which were
accepted after the tests mentioned above.
Our definition of a candidate calibrator is provided in
Sect.~\ref{transfun}, and more details on the
data statistics are given in Sect.~\ref{observations}.

\section{The interferometric transfer function}\label{transfun}
The squared visibility of a source with angular diameter $\phi$,
observed with an interferometer having a baseline $B$, at the
wavelength $\lambda$, is defined as
$$
V^2= \left[{2{\rm J}_1(\pi \phi B / \lambda) \over 
\pi \phi B / \lambda }\right]^2
$$
Here, and throughout the paper, we assume always a uniform
stellar disc. The effect of limb-darkening
can be investigated by long-baseline interferometry
(Wittkowski et al. \cite{Witt01}, \cite{Witt04}), but it is relatively
small and can be neglected for the purpose of this paper.
Due to degradations introduced by the atmosphere, as well
as by the optics and mechanics of the interferometer, the actual
squared visibility observed is $V^2_{\rm o}$, with
$V^2_{\rm o} < V^2$. The transfer function TF is defined as the
ratio
$$ {\rm TF} = V^2_{\rm o} / V^2$$
In order to perform scientific measurements on stars with an
unknown angular diameter, it is necessary to estimate TF
from stars for which the angular diameter is known. This is in
essence the process of calibration of an interferometer. Since
the TF is subject to the influence of atmospheric turbulence, it
must be measured as frequently as possible. It is also desirable
to estimate it from stars with a similar elevation, and ideally
also in a similar direction to the scientific target stars.

It can be assumed that the process of determining an interferometric
transfer function and the guidelines for its validity could
be similar to those established for example in studies of
speckle interferometry (Aitken \cite{Aitken}, Leinert \cite{Lei94}). 
We note however that some differences
could exist, for example linked to the different atmospheric properties
over distances significantly longer than the outer turbulence
scale, and this effect could be site-dependent
(Abahamid et al. \cite{Abahamid}).  There is no
general study of the regime of validity of transfer
function determinations for long-baseline interferometry.

For the purpose of this paper, we have used the whole database
of VINCI measurements after automated processing by the pipeline,
using only OBs which were not flagged for rejection as explained in
Sect.~\ref{vinci}. We have identified a number of stars
which satisfied the main criteria to be considered as candidate
calibrators.
These include the stability, the absence of nearby companions, 
and suitable spectral types. More details 
are given by Percheron et al. (\cite{Per03}). 
A summary of the
observations satisfying these criteria is given in the
last column of Table~\ref{table_summary}.

Essentially, we have used as the main
source the VLTI Calibrator catalog available to users
of the ESO web tools. This in turn is based on entries from
CHARM2 (Richichi et al. \cite{CHARM2}), from the list
of MIDI calibrators (T. Verhoelst, priv. comm.) and
from the catalog of Bord{\'e} et al. (\cite{Bor02}). Very few
stars come from other miscellaneous sources.
Details are given in the on-line material 
(see Table~4).
Our
list includes stars which cannot always  be
considered good calibrators in a strict sense.
In particular,
since MIDI is an instrument working in the 10$\mu$m
range, the compilers of its list of calibrators
were less inclined to reject stars with some
known variability in the near-IR, as long as
it did not affect its mid-IR behaviour.
As a result, some of the MIDI calibrators have
quite large angular diameters and/or are variable
in the near-IR. We do not concern ourselves here with
this issue, and we postpone a critical selection
to the next paper.

On the basis of the available angular diameters, 
we have computed the TF$_{i,j}$ for each calibrator $j$ 
and for each night $i$, and the associated error
$\Delta$TF$_{i,j}$. We have then computed for each night the
weighted average transfer function and its error
$$ {\rm TF}_{i} = {\sum_j \Delta{\rm TF}_{i,j}^{-2} \cdot {\rm TF}_{i,j} \over
\sum_j \Delta{\rm TF}_{i,j}^{-2}} $$
$$ \Delta{\rm TF}_{i} = \left(
\sum_j \Delta{\rm TF}_{i,j}^{-2} \right)^{{1\over2}} $$

We stress that this is a rather rough approximation,
since it is based on a pre-defined set of $\phi_j$ values and
we know that many of the listed calibrator diameters are not
accurate to the precision demanded by the VLTI/VINCI measurements.
The values we have
used are those provided in the references mentioned
above, and in many cases the diameters are either for a
wavelength other than the K band, or 
several determinations are available.
We have not attempted to derive a definitive value
for each star, but have taken instead what we considered
a representative value.
Nevertheless, we estimate that the
derived TF as a function of time is a sufficient approximation
for a statistical discussion as provided in Sect.~\ref{observations}.
It will be the subject of a second
paper 
to go into more detail for a selected
sample of calibrators, and in particular to provide refined
angular diameter values.
Note that the error terms $\Delta$TF$_{i,j}$ include also the uncertainties
on the angular diameters $\Delta\phi_j$, which at this stage
are necessarily as tentative as the initial values of
angular diameters themselves.

\section{Statistics of the observations}\label{observations}
The VLTI started operations in the middle of March 2001, using the
VINCI instrument and the SID. 
Since then, observations have continued
almost uninterrupted on every useful night.
The main goal of this long period of observations has been
the commissioning of the VLTI, including the characterization
of several baselines (see Fig.~\ref{fig_layout}), 
of the delay lines, of the several subsystems,
and of the complex software overviewing
the whole system
(Sch{\"o}ller et al. \cite{schoeller}).
More recently, scientific observations have started,
including the so-called Science Demonstration (SD), Guaranteed
Time Observations (GTO) for MIDI, and Open Time (OT) proposals
by the community.
All commissioning and SD
data taken on the sky with possible scientific relevance
are being released to the public on a regular basis. 

These public data form the basis of the material discussed in this
work.  We have considered observations obtained from
the night of first fringes, 16 March 2001, to
22 July 2004, which corresponds to the
latest public release.
This makes a total of 1225 nights.
The actual number of nights with data 
useful for our reprocessing is about 60\% of this,
as detailed in 
Table~\ref{table_summary}. The rest of the time was
spent mostly on integration and commissioning
activities with VINCI as well with
other instruments, not on the sky, or for GTO and OT
scientific observations 
which are protected by proprietary rules.
Overall, the fraction of time
in which the VLTI was idle from observations or technical
work was less than 2 months over the first 3.4 years, which represents
in itself an impressive performance.

Some general statistics are presented in 
Table~\ref{table_summary}.
The total volume of data processed 
amounted to 20849 OBs,
or $\approx$150~Gb of raw FITS files.
Of these,
18402 were successfully processed by the pipeline.
Note a slight difference to the total number provided in 
Table~\ref{table_summary}, because for 50 OBs the baseline
information was not reliable.
After the selections
described in Sect.~\ref{vinci}, 15123 OBs were accepted as
having a satisfactory quality, for a total of 287 stars.
A significant fraction of these stars were actually
science targets, or otherwise considered unsuitable as
calibrators.
After removing these entries, our database
includes 12066 OBs of good quality on 191 candidate
calibrators. 
Details are given in Table~4, which
is provided as on-line material.
In particular, we provide the following: a list of the candidate
calibrators; their main parameters such as coordinates,
magnitudes, spectral types; the baselines on which they
were observed and how many times.

\begin{figure}
\resizebox{\hsize}{!}{\includegraphics{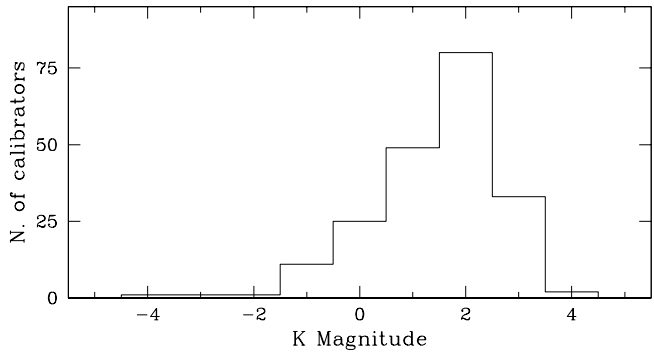}}
\caption[ ]{K magnitude distribution of the observed candidate
calibrators.
}\label{dist_mag}
\end{figure}

\begin{figure}
\resizebox{\hsize}{!}{\includegraphics{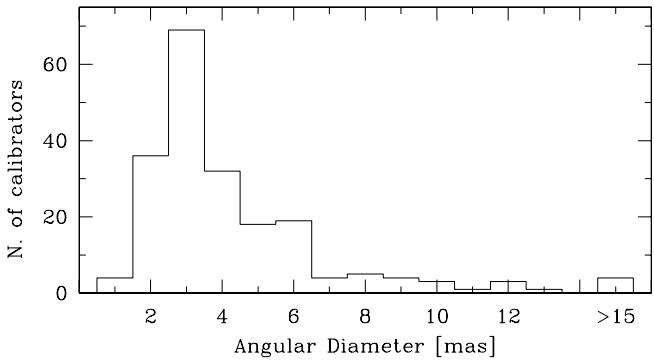}}
\caption[ ]{Angular diameter distribution of the observed candidate
calibrators.
}\label{dist_diam}
\end{figure}
Figs.~\ref{dist_mag} and \ref{dist_diam} show the magnitude
and brightness distributions of the observed candidate calibrators. 
The K magnitudes
have been obtained following the same criteria and using the
same information as described in Richichi et al. (\cite{CHARM2}).
Note that 
nominal values of the diameters are used, as stressed 
in Sect.~\ref{transfun}. It can be seen that most of the sources
are within the brightness limits of K=0 and K=3\,mag. This is a
consequence of the typical brightness limit attained with VINCI
on the small SID telescopes.
The detector integration times used with VINCI
during the commissioning observations varied between
0.1 and 20\,ms, however the most commonly used
values (accounting for 80\% of the data)
were between 0.8 and 1.0\,ms. These should be
understood as time spent on a scan.
When the total number of scans in an OB is
considered, the corresponding typical integration times
were about 0.4 to 0.45\,s.

Also, it can be noted from Table~\ref{table_summary}
that the majority of the observations were carried out
on SID telescopes, with the UTs accounting for only
3.0\% of the total number of accepted OBs.
We stress that for the present work only 
observations on UTs without MACAO were used.
Typically, there is a difference of about $\la$4-5\,mag between
the performance of VINCI at the UTs  without MACAO and at the SID.
However, the VLTI Calibrator catalog 
was based on the SID performance and this explains
why Fig.~\ref{dist_mag} does not extend to
K$\approx$10 which would be a more realistic limit
for the UTs.

The angular diameters peak at 3\,mas, providing a good coverage
of near-IR calibrators especially for baselines up to about
100\,m. For longer baselines, this set of calibrators can be
used but most of them will be significantly resolved and the
accuracy of calibration could be affected. For longer wavelengths,
such as in the 10$\mu$m range, all VLTI baselines can be adequately
calibrated. It should be noted that a small number of
candidate calibrators have very large angular diameters:
these are mainly from the MIDI list, and are unsuitable
for near-IR  calibration.

\begin{figure}
\resizebox{\hsize}{!}{\includegraphics{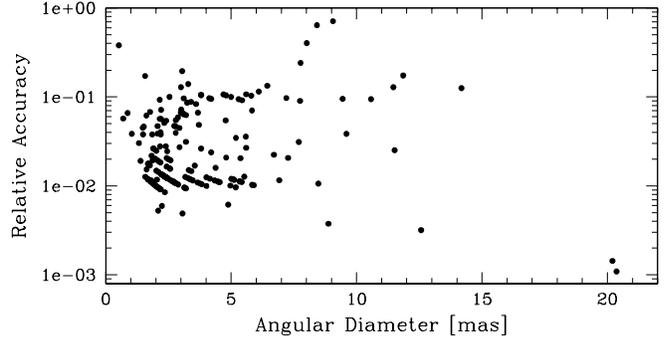}}
\caption[ ]{Diameter accuracy distribution of the observed candidate
calibrators.
}\label{dist_acc}
\end{figure}

Ideally, the best calibration accuracy is achieved when the
calibrator candidate can be assumed to be effectively unresolved
at the given wavelength and baseline. In practice, as just mentioned
in the context of Fig.~\ref{dist_diam}, calibrators can be
marginally resolved. In this case, the relative accuracy with which
the angular diameter is known will affect directly the accuracy
of the final calibrated visibility of the science target.
Fig.~\ref{dist_acc} shows that about 70\% of the observed candidate
calibrators have angular diameters known, or presumed, with
better than 5\% accuracy.
In this figure, it can be noted how some of the points with
higher accuracy and smaller angular diameters seem to be lined
up along a few precise curves. These are candidate calibrators
for which the angular diameter has been determined indirectly,
such as those present in the list of
Cohen et al.  (\cite{Coh99}). The curves are the consequence
of a precise link between the angular size and the accuracy
of the determination, and are probably produced by the
process of modelling and the associated parameter uncertainties.

\begin{figure}
\resizebox{\hsize}{!}{\includegraphics{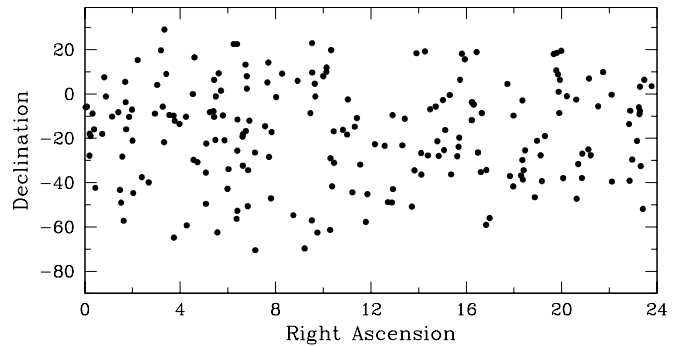}}
\caption[ ]{Sky distribution of the observed candidate
calibrators.
}\label{dist_sky}
\end{figure}

It can be appreciated from Fig.~\ref{dist_sky} how the
sky distribution
coverage is relatively uniform. For each of the 191 candidate
calibrators, the nearest one is on average 
$5\fdg8$, and never more than
$15\fdg8$, away. More important is the distance to the
nearest calibrator for any point in the sky. Restricting
the declination range between $-85\degr$ and $+35\degr$,
equivalent to an airmass of 2 as seen from Paranal,
the average distance to the nearest calibrator is
$9\fdg0$, and it never exceeds $30\degr$.
If the range of
declinations is restricted so as to limit the airmass to no more
than 1.5, the 
average distance to the nearest calibrator is
$7\fdg0$, and it never exceeds $26\degr$.

As mentioned in Sect.~\ref{transfun}, unfortunately 
no precise studies are available on the dependence of the accuracy
of calibration from the angular distance between science
and calibrator sources. It is generally assumed that such
a distance should be less than $\approx15\degr$ for reliable
results, and it is probably desirable to keep it even smaller.
Davis et al. (\cite{Davis}) found almost
no changes in the VLTI transfer function over a wide
range of zenith angle, although this study was
admittedly limited to one night only.
From this point of view, it appears that the coverage
provided by the list of candidate calibrators 
observed by the VLTI is sufficient for most applications.

\section{The VLTI Transfer Function}\label{vltitf}
\begin{figure}
\resizebox{\hsize}{!}{\includegraphics{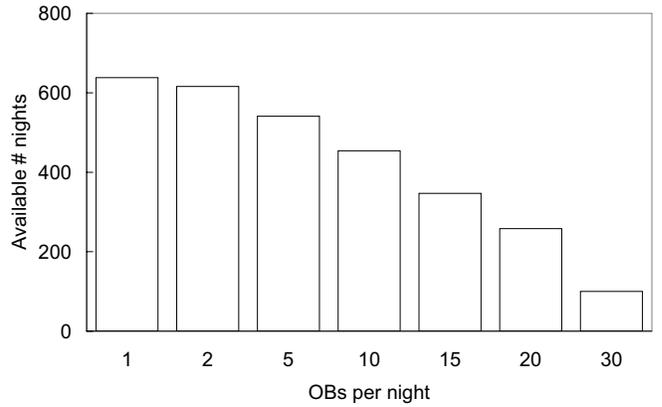}}
\caption[ ]{
Histogram of the number of nights according to the
number of OBs (observation blocks) carried out
on candidate calibrators. The maximum number of
OBs ever observed in one night was 97.
}\label{fig_nightob}
\end{figure}

One main result that can be drawn from the large
database of VLTI observations of candidate calibrators
is the evaluation of the VLTI transfer function, its
accuracy and its stability, both on a nightly basis
and its long-term behaviour. For this, we have
computed the transfer function TF according
to the definitions given in Sect.~\ref{transfun}.
We stress again that our computations are based
on a set of angular diameters, and their uncertainties,
which at this stage are only first approximations.
Given the large volume of observations and the number of candidate
calibrators, we should not be affected by systematic
biases in our conclusions. In order to further reduce
this risk, we have chosen to use only nights in which
at least 23 OBs were available. This number has been
chosen as a compromise between having a good number
of useful nights (about 200 according to this choice),
and keeping at the same time only nights with a very
high number of observations. A histogram of the
number of nights as a function of OBs observed is
shown in Fig.~\ref{fig_nightob}.

\begin{figure}
\resizebox{\hsize}{!}{\includegraphics{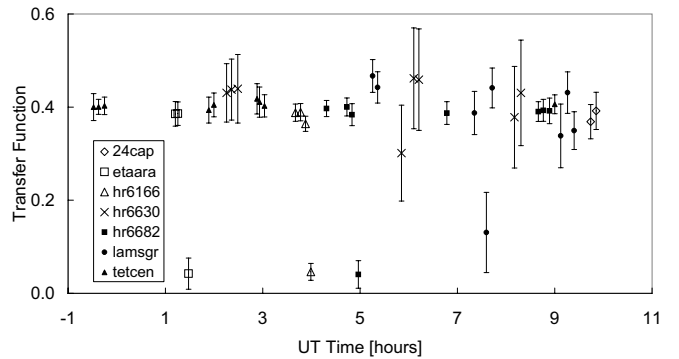}}
\caption[ ]{
Calibrators observations on the night of 2 June 2002,
using the E0-G1 baseline. See the text for a discussion
on the rejection of discrepant visibility points.
}\label{fig_2jun02}
\end{figure}

Additionally, we have chosen to discard from each night
those OBs which produced a TF value departing by more
than 3 standard deviations ($\sigma$) from the night average.
Fig.~\ref{fig_2jun02} shows an example in which
three OBs recorded between 1 and 5 hours UT
are obviously discrepant and were discarded according
to the above criterium. It can be noted 
that another OB appears to be discrepant, around 7.5 hours UT,
but it was not discarded because it did not violate the
3$\sigma$ criterium.

\begin{figure}
\resizebox{\hsize}{!}{\includegraphics{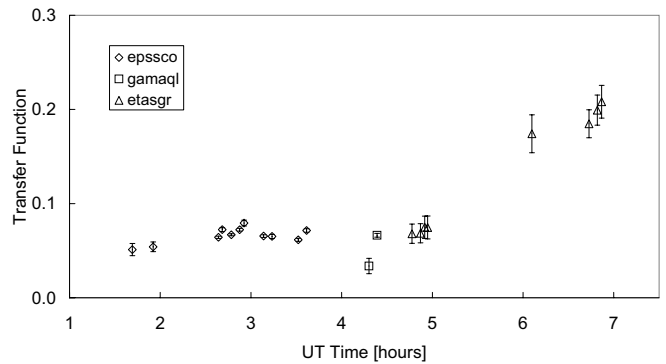}}
\caption[ ]{
Same as Fig.~\ref{fig_2jun02}, for the night of
23 July 2001. The baseline used was E0-G0.
No measurements were rejected, as explained in the text.
}\label{fig_23jul01}
\end{figure}

On the other hand, in spite of
this simple rejection criteria, the remaining validated
OBs are not necessarily always satisfactory. An example
is shown in Fig.~\ref{fig_23jul01}, where intuitively
one would be led to consider the OBs recorded after
6 hours UT as discrepant. However, they do not differ
by more than 3$\sigma$ from the night average. Also,
it is difficult to find an obvious reason for this
apparent discrepancy, since they are related to a
star ($\eta$~Sgr) for which several apparently good results
were obtained earlier in the night,
and no
obvious correlation is seen with ambient parameters
such as seeing, wind, coherence time.

We did not push our analysis to
identify specific reasons for each rejection,
or for the inclusion of apparently discrepant points.
As stated earlier, our goal is mainly to derive
global statistical properties, and we will devote
another paper to the individual evaluation of
selected results and to a refined list of
angular diameters of selected calibrators.
In particular, as mentioned in Sect.~\ref{transfun},
some of the MIDI calibrators are known
to be unsuitable for the near-IR.

\begin{figure}
\resizebox{\hsize}{!}{\includegraphics{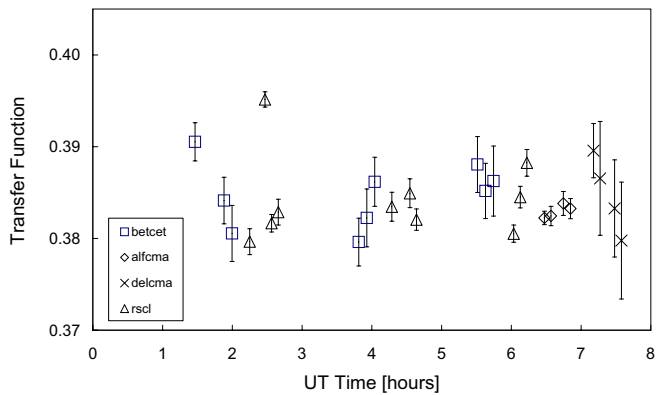}}
\caption[ ]{
Calibrators observations on the night of 14 November 2001,
using the E0-G0 baseline. The weighted average transfer function
is 0.384
with a relative error of only 0.5\%. This can be considered as
an example of a good night at the VLTI. 
}\label{fig_gnight}
\end{figure}

\begin{figure}
\resizebox{\hsize}{!}{\includegraphics{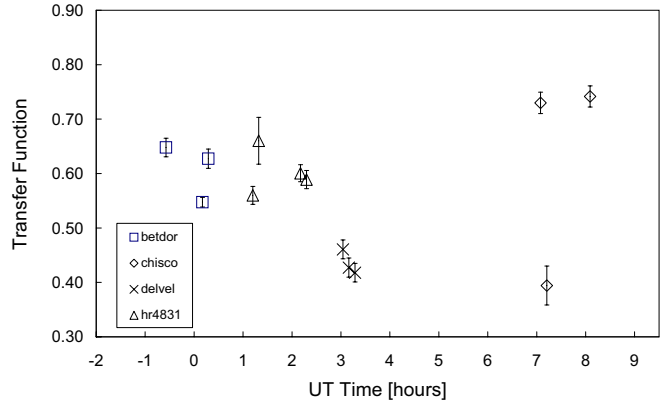}}
\caption[ ]{
Calibrators observations on the night of 20 April 2003,
using the B3-M0 baseline. The weighted average transfer function
is 0.574
with a large relative error of 15.8\%. This can be considered as
an example of a very bad night at the VLTI. 
Note the large scatter in the measurements
of each star.
}\label{fig_bnight}
\end{figure}

Figs.~\ref{fig_gnight}-\ref{fig_bnight} show examples
of two nights of calibrator observations, and the corresponding
TF average value. It can be appreciated how, depending also
on ambient conditions and in particular on the atmospheric
seeing and coherence time, the observations might show different
levels of scatter and quality.
On the night of November 14, 2001, the seeing was generally
constant at the level of $0\farcs6$, increasing towards
$1\farcs0$ only at the end of the night. The coherence time
was also approximately constant around 4\,ms. Note that
these values are computed at visual wavelengths, and at a location
on the mountain top separate from the VLTI telescopes. 
On the night of April 20, 2003 on the contrary,
the seeing fluctuations were more pronounced, with values
between $0\farcs3$ and $1\farcs1$ and increasing to over 
$1\farcs5$ at the end of the night. Also the coherence time 
changed significantly during the night, ranging between
2 and 8\,ms.

\begin{figure*}
\resizebox{\hsize}{!}{\includegraphics{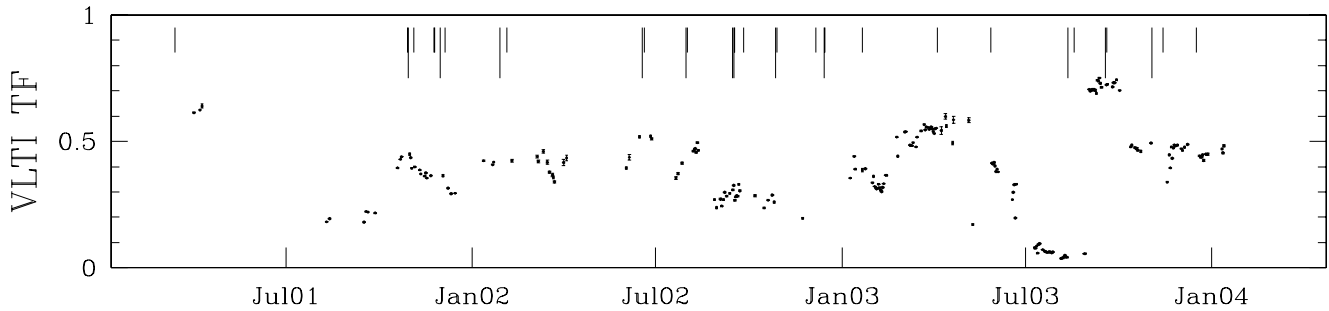}}
\caption[ ]{Long term plot of the night-average VLTI transfer function, as
computed from the nominal diameters of the candidate
calibrators. Changes of baseline are marked by the vertical 
segments at the top (short for SID, long for
UT baselines). At this scale, the error bars are barely visible.
}\label{dist_tf}
\end{figure*}

The overall TF of the VLTI, computed under the assumptions
illustrated so far, is shown in Fig.~\ref{dist_tf} for the
totality of the available data.
Firstly, one can notice a long period, in the summer of 2001, in
which almost no transfer function is available. Indeed, in this period
sky observations at the VLTI were almost completely halted, due
to a problem in the MONA beam combiner. The problem, caused by
temperature-dependent polarization of the single-mode fibers,
recurred several times in the lifetime of this
beam combiner. It required frequent adjustments of the
local temperature  of the fibers, and it became particularly
worrysome in the summer of 2003, when a period of very
low values of the transfer functions can be noticed.
It should be stressed that this instability of the interferometric
efficiency of the MONA beam combiner is characterized by relatively
long time scales, of the order of weeks. As such, it is quite
evident in the plot of Fig.~\ref{dist_tf}, however it did not
affect significantly observations collected during the same night.

A second consideration is that  the VLTI transfer function
varied at each baseline change, but not dramatically. This is
due to the fact that such changes can involve movements and
swaps of one or more relay mirrors in the delay line tunnel.
In particular, it can be noted that there are no major systematic
changes of the transfer function between SID and UT baselines.
The biggest changes in the transfer function can be easily associated
with temperature adjustments in the MONA beam combiner, realignments of
the general optics, and similar intervertions. When a baseline
was in place for a long period of time, long-term drifts are
evident, which can also be related to manual interventions
on the components of the VLTI.

The smoothest part of the long-term transfer function is
between October 19, 2001 and May 6, 2003. 
Before and after
this range, the MONA problems mentioned above altered
the long-term transfer function significantly.
For the 115 nights inside this period,
each having 
more than 23 OBs on calibrators,
the average transfer function was $0.405\pm0.034$.
Of course, it should be kept in mind that this relates
to the long-term behaviour only. On a nightly basis
the relative accuracies ranged from 0.1\% to 2.9\%,
with an average of 0.68\% which can be considered
as a satisfactory result.

\section{Conclusions}
We have presented and discussed a large set of
interferometric observations of candidate calibrators
obtained with the VLTI, equipped with the VINCI/MONA
beam combiner, over a period of more than three years. 
The observations have been obtained
in the framework of the VLTI commissioning, and
are publicly available from the ESO web site.
The data have been processed by an automated
pipeline, and have been subjected to a
number of quality criteria. 
A subset of 12066 observations
has been selected. 

From this, we have computed the VLTI transfer function
and discussed its main statistical properties both
on a nightly basis and in the long term.
The typical transfer function accuracy (in the squared visibility
sense) of the VLTI with VINCI/MONA 
of order of 0.7\% on a nightly basis, on a sample of 115 nights,
each having at least 23 observations of candidate calibrators.
For the long term behaviour, the transfer 
function shows marked fluctuations which generally can be 
explained in terms of slow degradations of the MONA
beam combiner, and to a lesser extent in terms of baseline
changes, realignments, and similar interventions. 
From October 2001 to May 2003, the average value
of the transfer function was 0.405$\pm$0.034.

We present as on-line material a list of 191 candidate
calibrators, together with their main characteristics
and the statistics of their observations at the VLTI.
This list is preliminary, in the sense that some of these
sources, extracted from a catalog of calibrators for
the mid-infrared beam combiner MIDI, are not ideally suited for 
calibration in the near-infrared. Also the angular diameter
values are tentative, having been extracted from the
available literature without any effort to perform
a consistency check against the VLTI observations.
Another paper is in preparation, with the goal of deriving
a new set of angular diameters for a restricted sample
of validated calibrators, by means of a global solution
to the VLTI observations presented here.

\begin{acknowledgements}
The VLT Interferometer is the result of more than a decade of 
work in planning
and engineering, and of several intense years
of efforts on the site.
We would like to recognize the many tens of 
people involved in
this enterprise, even without being able to acknowledge each
by name.
This research has made use of the \textit{Simbad} database, operated at the
Centre de Donn{\'e}es Astronomiques de Strasbourg (CDS), and of
NASA's Astrophysics Data System Bibliographic Services (ADS).
\end{acknowledgements}

\end{document}